\documentclass[fleqn,twoside]{article}
\usepackage{espcrc2}

\usepackage{graphicx}
\usepackage[figuresright]{rotating}

\newcommand{\AmS}{{\protect\the\textfont2
  A\kern-.1667em\lower.5ex\hbox{M}\kern-.125emS}}

\title{
SU(3) Predictions of $B\to PP$ Decays in the Standard Model}

\author{H.-K. Fu, X.-G. He, Y.-K. Hsiao and J.-Q. Shi \address{
Department of Physics, National Taiwan University,
Taipei, Taiwan}
        \thanks{
This work was supported in part by ROC National 
Science Council under grant NSC 90-2112-M-002-014, and in part by the
Ministry of Education Academic Excellence Project 89-N-FA01-1-4-3.}}
 
\begin{document}

\begin{abstract}
With SU(3) symmetry one only needs 13 hadronic
parameters to describe $B\to PP$ decays
in the Standard Model.  When annihilation
contributions are neglected, only 7 hadronic parameters are
needed. These parameters can be determined from existing experimental
data and some 
unmeasured branching ratios and CP asymmetries of the type $B\to PP$ 
can be predicted.
In this talk we present SU(3) predictions of branching ratios and 
CP asymmetries for $B\to PP$ decays in the Standard Model.
\vspace{1pc}
\end{abstract}

\maketitle

\section{SU(3) Parameters for $B\to PP$}

In Standard Model (SM) the decay amplitudes 
for $B$ meson to two pseudoscalar SU(3)
octet mesons, $B \to PP$, have both tree and penguin
contributions and can be written as\cite{0}
\begin{eqnarray}
&&A(B\to PP) = <PP|H_{eff}^q|B>\nonumber\\
&&= {G_F\over \sqrt{2}}[V_{ub}V^*_{uq} T(q)
+ V_{tb}V^*_{tq}P(q)].
\end{eqnarray}
As far as the SU(3) structure is concerned, the effective Hamiltonian
contains $\bar 3$, $6$, and $\overline{15}$ representations of SU(3) 
which define several invariant amplitudes\cite{1},

\begin{eqnarray}
&&T= A_{\bar 3}^TB_i (\bar 3)^i (M_l^k M_k^l) + C^T_{\bar 3}
B_i M^i_kM^k_j(\bar 3)^j \nonumber\\
&&+ A^T_{6}B_i (6)^{ij}_k M^l_jM^k_l + C^T_{6}B_iM^i_j(6
)^{jk}_lM^l_k\nonumber\\
&&+A^T_{\overline {15}}B_i (\overline {15})^{ij}_k M^l_jM^k_l +
C^T_{\overline
{15}}B_iM^i_j
(\overline {15} )^{jk}_lM^l_k\;,\nonumber
\label{tpp}
\end{eqnarray}
where $B_i = (B_u,\;B_d,\;B_s)$ and $M$ is the SU(3) pseudoscalar
octet.
$C_6-A_6$ always appear together. We use $C_6$ only.

In general there are both tree and penguin amplitudes 
$C(A)^{T,P}_{\bar 3,6,\overline{15}}$. 
In the SM the amplitudes
$C(A)^T_{6,\overline{15}}$ and $C(A)^P_{6,\overline{15}}$
are related by the Wilson coefficients $c_i$ of the relevant
operators\cite{2,3},

\begin{eqnarray}
{C^P_6\over C^T_6} &=&
- {3\over 2}
{c_9^{tc} - c_{10}^{tc}\over
c_1-c_2-3(c_9^{uc}-c_{10}^{uc})/2},\nonumber\\
{C^P_{\overline {15}}(A^P_{\overline {15}})
\over 
C^T_{\overline {15}} (A^T_{\overline {15}})}
&=& -{3\over 2}
{c_9^{tc}+c_{10}^{tc}\over c_1+c_2-3(c^{uc}_9+c_{10}^{uc})/2}
.
\label{P2T}
\end{eqnarray}
Without loss of generality, one can set $C^P_{\bar 3}$ to be real. One 
only needs to use 13 real independent parameters to describe $B\to PP$ in the
SM which we choose to be\cite{2,3}

\begin{eqnarray}
&&C_{\bar 3}^P,\;\;C_{\bar 3}^T e^{i\delta_{\bar 3}},\;\;
C^{T}_6e^{i\delta_6},\;\;
C^{T}_{\overline{15}}e^{i\delta_{\overline{15}}},\nonumber\\
&&A^T_{\bar 3}e^{i\delta_{A^T_{\bar 3}}},\;\;
A^P_{\bar 3} e^{i\delta_{A^P_{\bar 3}}},\;\;
A^T_{\overline{15}} e^{i\delta_{A^T_{\overline{15}}}}.
\end{eqnarray}
The phases are defined such that all $C_i^{T,P}$
are real positive numbers.
The amplitudes $A_i$ correspond to annihilation contributions and
are expected to be small. If they are neglected, there are
only 7 independent hadronic parameters.

Using available data on, $B\to K \pi, \pi\pi, KK$, one can obtain 
information about the hadronic parameters and predict other
decay branching ratios and CP violating asymmetries.
We carried out a $\chi^2$ analysis\cite{4} assuming SU(3) symmetry 
with the $B\to PP$ data\cite{5} 
shown in 
Table 1. In our numerical analysis the KM matrix elements $V_{us} 
= \lambda$,
$V_{cb} = A\lambda^2$, $V_{ub} = |V_{ub}| exp(-i\gamma)$ fixed
by $\lambda = 0.2196$, $A = 0.835$ and 
$|V_{ub}| = 0.09|V_{cb}|$\cite{6} and $\gamma = 59^\circ$ as determined from
other data\cite{4}. 
We study both the cases with and without annihilation
contributions and compare the results.

\begin{table}[htb]
\caption{Experimental data on the branching ratios and CP asymmetries 
obtained by averaging different data with the assumption that they obey 
uncorrelated Gaussian distribution. 
The branching ratios for $B\to PP$ are shown in units of $10^{-6}$.}\label{br}
\begin{center}
\begin{tabular}{|l|l|}
\hline
 Branching ratio and  &Averaged \\
 CP asymmetries       & Value \\ \hline \hline
$Br(B_d\to \pi^+\pi^-)$   &            $5.2\pm0.6$\\  \hline
$Br(B_u\to \pi^-\pi^0)$  & $4.9\pm1.1$\\         \hline
$Br(B_d\to K^+\pi^-)$    &  $18.6 \pm 1.1$\\   \hline
$Br(B_u\to K^-\pi^0)$    &       $11.5\pm 1.3$\\ \hline
$Br(B_u\to \bar K^0 \pi^-)$  &         $17.9\pm 1.7$\\   \hline
$Br(B_d\to \bar K^0\pi^0)$   &        $8.8\pm 2.3$\\ \hline
$Br(B_u\to K^-K^0)$          &         $0\pm 0.8$\\   \hline
$Br(B_d\to\bar K^0 K^0)$     &         $1.8\pm2.5$\\ \hline
$Br(B_d\to\pi^0\pi^0)$       &         $1.7\pm 0.9$ \\        \hline
$Br(B_d\to K^+K^-)$     &         $0\pm0.3$\\   \hline
\hline
$A^{B_d}_{ K^-\pi^0}$    &       $-0.05\pm0.09$\\    \hline
$A^{B_d}_{K^-\pi^+}$    &       $-0.05\pm0.05$\\    \hline
$A^{B_u}_{\bar K ^0\pi^-}$     &      $0.04\pm0.08$\\  \hline
$A^{B_d}_{\pi^+\pi^-}$  &        $0.42\pm0.22$ \\  \hline
$A^{B_u}_{\pi^-\pi^0}$    &   $0.13\pm0.21$\\ \hline
\end{tabular}
\end{center}
\end{table}

\section{Without Annihilation Contributions}

For the case without annihilation contributions, there are
only 7 hadronic parameters. The results for the best fit values and
their 1$\sigma$ ranges are shown in Table 2.

\begin{table}
\caption{The best fit values and their $1\sigma$ errors of the hadronic parameters
using all data in Table \ref{br} with 
annihilation terms set to be zero.}\label{fp}
\begin{center}
\begin{tabular}{|c|c|c|}
\hline
            &   best value   &   error \\  \hline
$   C^P_{\bar 3}    $   &   0.138   &   0.003  \\  \hline
$   C^T_{\bar 3}    $   &   0.248   &   0.111  \\  \hline
$   C^T_6   $   &   0.155   &   0.112      \\  \hline
$   C^T_{\overline{15}} $   &   0.142   &   0.014   \\  \hline
$   \delta_{  \bar 3}   $   &   $38.10^0$   &   $29.69^0$    \\  \hline
$   \delta_ 6   $   &   $83.17^0$    &   $35.97^0$  \\  \hline
$   \delta_{\overline{15}}  $   &   $4.78^0$    &   $17.84^0$ \\  \hline
\end{tabular}
\end{center}
\end{table}

Using the above determined hadronic parameters, 
one can easily obtain the branching ratios and 
CP asymmetries for other $B\to PP$. 
We used the following definition for the CP violating rate asymmetry,
\begin{eqnarray}
A^{B_i}_{PP} = {\Gamma( B_i \to PP) - \Gamma(\bar B_i \to \bar P \bar P)\over
\Gamma(B_i\to PP) + \Gamma(\bar B_i\to \bar P\bar P)}.
\end{eqnarray}

In general $P$ can be any one of 
the SU(3) pseudoscalar octet mesons, $\pi$, $K$ and $\eta_8$. 
Here we will limit our study 
to $P = \pi, K$ to avoid complications associated with $\eta_1$ and $\eta_8$ mixings. 
In this case there are total 16 decay modes. Among them
the decay amplitudes for $B_d \to K^-K^+$, $B_s \to \pi^-\pi^+, \pi^0\pi^0$
only receive annihilation contributions.  
Since we have neglected annihilation contributions they 
would have vanishing branching ratios.
At present none of them have been measured experimentally. The present bound on
$B_d \to K^- K^+$ is consistent with this prediction.

The predictions for
the branching ratios of $B_s \to K^+ \pi^-, K^0 \pi^0, K^-K^+, K^0 \bar K^0$ 
decays are shown in Table 3. 
These decay modes are predicted to be large and 
can be measured at hadron colliders. 
The standard model and SU(3) flavor symmetry can be tested.

The CP asymmetries for some of the decays are shown in Table 4. 
In the SU(3) limit there are some relations between rate differences
defined as,
$\Delta^{B_i}_{PP} = \Gamma(B_i \to PP) - \Gamma(\bar B_i \to \bar P \bar P)$,
between $\Delta S = 0$ and $\Delta S = -1$ modes due to 
a unique feature of the SM in the KM matrix element that $Im(V_{ub}V_{ud}^*V_{tb}^*V_{td})
= - Im(V_{ub}V_{us}^*V_{tb}^*V_{ts})$.
One has the following relations\cite{1,7}, 

\begin{eqnarray}
&&\Delta^{B_d}_{\pi^+ \pi^-} = \Delta^{B_s}_{K^+ \pi^-} = -\Delta^{B_d}_{
\pi^+ K^-},\nonumber\\
&&\Delta^{B_d}_{\pi^0 \pi^0} = \Delta^{B_s}_{K^- \pi^0} = -\Delta^{B_d}_{
\pi^0 \bar K^0}.
\end{eqnarray}

As can be seen from Table 4 that the best fit values for $A^{B_i}_{PP}$ 
can be large with several of them to be more than
30\%, such as the asymmetries for $B_s \to K^+ \pi^-, K^0 \pi^0$, and
$B_d \to \pi^0 \pi^0, \pi^+\pi^-$. The size of $A^{B_i}_{PP}$ for these
modes are  large, but can be easily understood from the fact that they all
have relatively small branching ratios.
Using the above relations, one would obtain

\begin{eqnarray}
&&A^{B_d}_{\pi^+\pi^-} = A^{B_s}_{K^+ \pi^-} 
\nonumber\\
&&= - A^{B_d}_{\pi^+ K^-} {Br(B_d \to \pi^+ K^-)
\over Br(B_d \to \pi^+ \pi^-)},\nonumber\\
&&A^{B_d}_{\pi^0\pi^0} = A^{B_s}_{K^0 \pi^0} 
\nonumber\\
&&= - A^{B_d}_{\pi^0\bar K^0} {Br(B_d \to \pi^0 \bar K^0)
\over Br(B_d \to \pi^0 \pi^0)}.
\end{eqnarray}
In all the above cases the ratios 
of the branching ratios are larger than one, a small
$A^{B_i}_{PP}$ of the decay mode on the right hand side can induce a 
large $A^{B_i}_{PP}$ for the decay modes on the left hand side.
These predictions can provide interesting tests for the SM.

\begin{table}[htb]
\caption{Predictions of branching ratios 
and $1\sigma$ ranges without annihilation terms in units of 
$10^{-6}$.}\label{BB}
\begin{center}
\begin{tabular}{|c|c|c|}
\hline
    &   best value   &   range \\  \hline
$B_s \to K^+ \pi^-$         &   4.8   &   (   5.3   ,   4.2   ) \\  \hline
$B_s \to K^0 \pi^0$         &   1.2    &   (   2.0    ,   0.7    ) \\  \hline
$B_s \to K^- K^+$       &   17.4 &   (   18.3   ,   16.5   )   \\  \hline
$B_s \to K^0 \bar K^0$      &   16.8  &   (   17.9   ,   15.8   )   \\  \hline
\end{tabular}
\end{center}
\end{table}

\begin{table}[htb]
\caption{Predictions of CP asymmetry without annihilation 
terms.}\label{AA}
\begin{center}
\begin{tabular}{|c|c|c|}
\hline
    &   best value   &   range       \\  \hline
$A^{B_u}_{K^- K^0}$        &   -0.09  &  (0.85, -0.91) \\  \hline
$A^{B_d}_{\pi^+ \pi^-}$        &    0.32   &   (0.46    , 0.18 )  \\  \hline
$A^{B_d}_{\pi^0 \pi^0}$         &   0.37  &   (   0.64   ,   -0.58)   \\  \hline
$A^{B_d}_{\bar K^0 K^0}$   &   -0.09  &   (   0.85  ,   -0.91  )    \\  \hline
$A^{B_u}_{\pi^- \bar K^0}$     &   0.00   &   (0.05  , -0.04 ) \\  \hline
$A^{B_u}_{\pi^0 K^-}$      &  -0.01  &   ( 0.06 ,  -0.10 ) \\  \hline
$A^{B_d}_{\pi^+ K^-}$      &   -0.09  &   ( -0.05 ,   -0.13 ) \\  \hline
$A^{B_d}_{\pi^0 \bar K^0}$     & -0.06 &   (0.06 ,   -0.13   ) \\  \hline
$A^{B_s}_{K^+ \pi^-}$      &   0.32   &   (   0.46   ,   0.18  )  \\  \hline
$A^{B_s}_{K^0 \pi^0}$      &   0.37  &   (   0.64   ,   -0.58  )  \\  \hline
$A^{B_s}_{ K^- K^+}$        &   -0.09  &   (   -0.05  ,   -0.13  ) \\  \hline
$A^{B_s}_{K^0 \bar K^0}$   &   0.00   &   (   0.05   ,   -0.04   ) \\  \hline
\end{tabular}
\end{center}
\end{table}

\section{With Annihilation Contributions}

In the analyses of the previous sections we have 
neglected annihilation contributions to 
$B\to PP$ decays. In this section we study the effects 
of the annihilation terms on $B\to PP$ decays. 
In this case we would have total 13 parameters. 
From Table \ref{br} we see that there are 15
experimental data points. In principle, the 13 hadronic parameters under consideration 
can be determined. In Tables \ref{ann1}, \ref{ann3} and \ref{ann2} we show the results
on the hadronic parameters, and some of the $B\to PP$ branching ratios 
and CP asymmetries.

\normalsize
\begin{table}[htb]
\caption{The best fit values and their errors for the hadronic parameters with 
annihilation terms.}\label{ann1}
\begin{center}
\begin{tabular}{|l|l|l|}
\hline
& best value& error \\ \hline
$   C^P_{\bar 3}    $   &   0.138   &   0.004 \\  \hline
$   C^T_{\bar 3}    $   &   0.208   &   0.181 \\  \hline
$   C^T_6   $   &   0.043   &   0.206    \\  \hline
$   C^T_{\overline{15}} $   &   0.141   &   0.014 \\  \hline
$   \delta_{  \bar 3}   $   &   $31.65^0$  &   $57.7^0$ \\  \hline
$   \delta_ 6   $   &   $97.74^0$   &   $147.97^0$ \\  \hline
$   \delta_{\overline{15}}  $   &   $8.54^0$   &   $21.67^0$ \\  \hline
$   A^P_{\bar 3}    $   &   0.025   &   0.042  \\  \hline
$   A^T_{\bar 3}    $   &   0.061   &   0.143   \\  \hline
$   A^T_{\overline {15}}    $   &   0.036   &   0.075  \\  \hline
$   \delta_{A^T_{\bar{3}}}  $   &   $73.46^0$   &   $107.18^0$ \\  \hline
$   \delta_{A^P_{\bar{3}}}  $   &   $-13.78^0$  &   $89.52^0$ \\  \hline
$   \delta_{A^T_{\overline{15}}}    $   &   $-131.12^0$  &   
$180.51^0$ \\  \hline
\end{tabular}
\end{center}
\end{table}

\footnotesize
\begin{table}
\caption{Predictions of branching ratios and $1\sigma$ range errors 
with annihilation 
terms in units of $10^{-6}$.}\label{ann3}
\begin{center}
\begin{tabular}{|l|l|l|}
\hline
    &   best value   &  range   \\  \hline
$B_s \to K^+ \pi^-$      &   4.1   &   (6.7  , 2.6      ) \\  \hline
$B_s \to K^0 \pi^0$      &   1.1   &   (2.2       ,0.4       ) \\  \hline
$B_s \to K^- K^+$        &   31.8  &   (51.9       ,7.1       ) \\  \hline
$B_s \to K^0 \bar K^0$   &   30.9  &   (50.7       ,6.5       ) \\  \hline
$B_s \to \pi^- \pi^+$    &   2.3   &   (9.5     , 0.0      ) \\  \hline
$B_s \to \pi^0 \pi^0$    &   1.1   &   (  4.7     , 0.0      ) \\ \hline
\end{tabular}
\end{center}
\end{table}

\scriptsize
\begin{table}[htb]
\caption{Predictions of CP asymmetry with 
annihilation terms.}\label{ann2}
\begin{center}
\begin{tabular}{|c|c|c|c|c|}
\hline
    &   best value   & error range \\  \hline
$A^{B_u}_{K^- K^0}$        &   -0.24  &   (   0.82   ,  -0.96  ) \\  \hline
$A^{B_d}_{\pi^0 \pi^0}$         &   0.19   &   (   0.72   ,   -0.99   ) \\  \hline
$A^{B_d}_{\bar K^0 K^0}$   &   0.78  &   (   1.00   ,   -1.00  ) \\  \hline
$A^{B_s}_{K^+ \pi^-}$      &   0.26   &   (   0.54   ,   0.03   ) \\  \hline
$A^{B_s}_{K^0 \pi^0}$      &   0.12   &   (   0.68   ,   -0.86   ) \\  \hline
$A^{B_d}_{\pi^0 \bar K^0}$     &   -0.02  &   (   0.13   ,   -0.12  ) \\  \hline
$A^{B_s}_{K^- K^+}$        &   -0.06  &   (   -0.02  ,   -0.24  ) \\  \hline
$A^{B_s}_{K^0 \bar K^0}$   &   -0.05   &   (   0.18   ,   -0.22  ) \\  \hline
\end{tabular}
\end{center}
\end{table}

\normalsize

From Table \ref{ann1} we see that the size of
the best fit annihilation parameters $A_i$ are small 
compared the  non-annihilation terms $C_{3,\overline{15}}$. 
This confirms the expectation that annihilation 
contributions are small. 
The allowed ranges are large and therefore can not 
rule out the possibility of having significant annihilation contributions. 
We have to wait improved experiments to obtain more precise information.

The branching ratios for  $B_d \to K^- K^+$, $B_s \to \pi^+\pi^-,
\pi^0\pi^0$ which only receive contribution from annihilation are not vanishing any more. 
The branching ratios are expected to be small. From Table \ref{ann3}, 
we indeed find that these branching ratios are smaller than others.  

It is interesting to note that although the 
annihilation amplitudes are small, in certain decay modes, such as 
$B_s \to K^+K^-$ and $B_s \to K^0 \bar
K^0$, the effects can be significant, the branching ratios
almost doubled. 
This is because that although  $A^P_{\bar 3}$ is small 
compared with $C_{3,\overline{15}}$, it is comparable with $C^T_{6}$, but
enhanced by a KM factor $|V_{tb}V_{ts}^*/V_{ub}V_{us}^*|$. 
These modes provide good places to 
study the annihilation contributions. 

\section{Discussions and Conclusions}

In this talk we have presented SU(3) predictions of branching ratios and 
CP asymmetries for some $B\to PP$ decays in the Standard Model. There can
be large CP violation in $B_d \to \pi^0\pi^0, \pi^+\pi^-$ and
$B_s \to K^0 \pi^0, K^- \pi^-$. Also several $B_s$ decays can have large 
branching ratios. We 
presented results obtained
with and without annihilation contributions. The results indicate that
the annihilation contributions are in general small, but can still
have large effect in several $B_s \to K^+ K^-, K^0 \bar K^0$ decay modes. 

In SM predictions for branching ratios and CP asymmetries
are possible is beacuse that 
SU(3) symmetry relates different decay modes. 
This symmetry is expected to be broken. In that
case more parameters are needed to describe the decays. 
The effects of SU(3) breaking have to be further studied.
We however expect that the general feature will not be be altered\cite{4} 
dramtically, and 
results obtained here can still provide some guidance
in searching for large CP asymmetries and branching ratios in
$B\to PP$ decays. Future experiments on $B\to PP$ 
will provide valuble information about the Standard Model.

\end{document}